 \documentclass[twocolumn,trackchanges]{aastex631}
\usepackage{CJK}

\received{XXX}
\revised{YYY}
\accepted{ZZZ}

\submitjournal{ApJ}

\shorttitle{Standing Shock in Low-Angular-Momentum Accretion}
\shortauthors{Mao et al.}
\graphicspath{{./}{figures/}}
\begin{document}
\begin{CJK*}{UTF8}{gbsn}

%\title{Low Angular Momentum Black Hole Accretion: A Test of Standing Shock Identification}  
\title{Low Angular Momentum Black Hole Accretion: First GRMHD Evidence of Standing Shocks}

\correspondingauthor{Jirong Mao}
\email{jirongmao@mail.ynao.ac.cn}

\author[0000-0002-7077-7195]{Jirong Mao}
  \affiliation{Yunnan Observatories,
    Chinese Academy of Sciences, 
    Kunming 650216, People's Republic of China}
  \affiliation{Center for Astronomical Mega-Science, 
    Chinese Academy of Sciences,
    Beijing 100012, People's Republic of China}
  \affiliation{Key Laboratory for the Structure and Evolution of Celestial Objects, 
    Chinese Academy of Sciences, 
    Kunming 650216, People's Republic of China}

\author[0000-0002-4064-0446]{Indu K. Dihingia}
  \affiliation{Tsung-Dao Lee Institute, Shanghai Jiao Tong University, 1 Lisuo Road, Shanghai 201210, People's Republic of China}
%  \affiliation{ShangHai Jiaotong Uin., }
%  \affiliation{Key Laboratory for the Structure and Evolution of Celestial Objects, 
%    Chinese Academy of Sciences, 
%    Kunming 650216, People's Republic of China}

\author[0000-0002-8131-6730]{Yosuke Mizuno}
  \affiliation{Tsung-Dao Lee Institute, Shanghai Jiao Tong University, 1 Lisuo Road, Shanghai 201210, People's Republic of China}
  \affiliation{School of Physics and Astronomy, Shanghai Jiao Tong University, 800 Dongchuan Road, Shanghai 200240, Peopleʼs Republic of China}
  \affiliation{Key Laboratory for Particle Physics, Astrophysics and Cosmology (MOE), Shanghai Key Laboratory for Particle Physics and Cosmology, Shanghai Jiao Tong University, 800 Dongchuan Road, Shanghai 200240, Peopleʼs Republic of China}
  \affiliation{Institut f\"{u}r Theoretische Physik, Goethe-Universit\"{a}t Frankfurt, Max-von-Laue-Strasse 1, D-60438 Frankfurt am Main, Germany}

\author[0000-0002-7025-284X]{Shigehiro Nagataki}
  \affiliation{Astrophysical Big Bang Laboratory (ABBL), RIKEN Pioneering Research Institute (PRI), 2-1 Hirosawa, Wako, Saitama 351-0198, Japan}
  \affiliation{RIKEN Center for Interdisciplinary Theoretical and Mathematical Sciences (iTHEMS), 2-1 Hirosawa, Wako, Saitama 351-0198, Japan} 
  \affiliation{Astrophysical Big Bang Group (ABBG), Okinawa Institute of Science and Technology Graduate University (OIST), 1919-1 Tancha,
Onna-son, Kunigami-gun, Okinawa 904-0495, Japan}



\begin{abstract}
%original sentences
%Black hole (BH) accretion is a crucial issue for the study of X-ray binaries (XRBs) and Active Galactic Nuclei (AGNs). Low angular momentum accretion is a special accretion mode for black hole (BH) accretion. 
%\textbf{Low angular momentum accretion as a special accretion mode is crucial for the study of black hole (BH) X-ray binaries (XRBs) and Active Galactic Nuclei (AGNs).}
Understanding the dynamics of low angular momentum accretion flow around black holes (BHs) is essential for probing extreme plasma behavior in strong gravity, where shock formation can naturally produce variability signatures. 
In this paper, we perform general relativistic magnetohydrodynamic (GRMHD) simulations of low angular momentum accretion flows onto a BH with different BH spins to investigate the accretion dynamics near the central BH region.
The simulation results show the standard and normal evolution (SANE) regime in all cases. In particular, we report the formation and persistence of standing shocks in low-angular-momentum accretion flows using multi (two and three)-dimensional GRMHD simulations for the first time. Previous studies did not detect such stable standing shock structures, making our findings a significant advancement in this field. The finding of shock dynamics
can be further associated with some radiation features, such as flares observed in Sgr~A$^\ast$ and quasi-periodic oscillation (QPO) signals detected in some XRBs and AGNs. 
\end{abstract}



\section{Introduction} \label{sec:intro}

Black hole (BH) accretion is one of the important physical processes in astrophysics. Accreting matter typically flows toward a central BH and forms an accretion disk \citep[e.g.,][]%{1973A&A....24..337S,1994ApJ...428L..13N,2024arXiv240406140D
{Frank_King_Raine_2002}. If the accretion flow has nearly no angular momentum, it creates quasi-spherical accretion flows that are called Bondi accretion \citep{1952MNRAS.112..195B}. On the other hand, by Keplerian angular momentum, %it could be 
a geometrically thin disk could be formed \citep{1973A&A....24..337S,2024arXiv240406140D}. Along with the angular momentum, the magnetic field also plays a crucial role in shaping various astrophysical phenomena, such as relativistic jets and disk winds \citep[e.g.,][]{2021MNRAS.505.3596D}.
Accretion flow can be
driven by magnetorotational instability (MRI), which introduces viscosity through %the 
turbulence %in the accretion flow 
\citep{1991ApJ...376..214B}. Differential rotation is necessary for developing the MRI. Depending on the net magnetic flux, the accretion might be in the standard and normal evolution (SANE) regime with a weak magnetic field \citep{2012MNRAS.426.3241N}. Alternatively, strong magnetic flux can be dragged to the central BH, and the magnetically arrested disk (MAD) is formed \citep{2003PASJ...55L..69N, 2012MNRAS.423.3083M}. In such a case, a collimated powerful outflow that is called a jet would be launched \citep{2011MNRAS.418L..79T}.

Low angular momentum accretion has been gaining attention recently to explain the complex observed behavior of Sgr~A$^*$ \citep{2018MNRAS.478.3544R,2023A&A...678A.141O,2024ApJ...967....4D}. Low angular momentum flows are known to have multi-transonic accretion solutions, where accretion flow changes its sonic state (subsonic to supersonic) multiple times while falling onto the black hole \citep[e.g.,][]{1987PASJ...39..309F,1980ApJ...240..271L,1981ApJ...246..314A}. Additionally, such multi-transonic flow could also make a transition from supersonic to subsonic through shock discontinuity if parameter ranges are appropriate \citep[e.g.,][]{1989ApJ...347..365C,1989PASJ...41.1145C,1992MNRAS.259..259N,1993PASJ...45..167N,1994MNRAS.270..871N}. There have been substantial studies in understanding and applying such solutions in different astrophysical contexts. For example, using semi-analytical approaches, a large number of parameter-space surveys have been performed to show that shock is ubiquitous for low-angular momentum flow but ceases to exist in extreme viscosity, radiative cooling, thermal conduction, magnetic fields, outflow, etc. \citep[e.g.,][]{2001ApJ...557..983D,2002ApJ...572..950T,2003A&A...409....1G,2006ApJ...645.1408T,Das2007,%Rajesh:2009vr,
2016ApJ...819..112L,Aktar-etal2017,Sarkar-etal2018,Dihingia-etal2018MNRAS,Dihingia-etal2020MNRAS,2023MNRAS.523.4431M, %2022MNRAS.516.3984R,
2022MNRAS.512.5771H,Sarkar-etal2023,Singh-etal2024,Jana-Das2024,Sarkar-etal2025}. There have been significant efforts towards numerical simulations to understand the formation of shocks in low-angular momentum flow in hydrodynamics and magnetohydrodynamics considering pseudo-Newtonian approximations \citep[e.g.,][]{Molteni-etal1994,Ryu-etal1995,Molteni-etal1996a,Molteni-etal1996b,Lanzafame-etal1998,2003ApJ...592..767P, Chakrabarti-etal2004,2008ApJ...689..391N,2009ApJ...696.2026N,2009ApJ...705.1503J,Giri-etal2010,2012NewA...17..254D,Okuda-Molteni2012, Okuda2014, Das-etal2014, Okuda-Das2015,Okuda-etal2019,Singh-etal2021,Okuda-etal2022,Debnath-etal2024,Huang-Singh2025}. These simulations suggest that the shock front can also exhibit oscillatory behavior. Such oscillatory shock could provide possible connections to the observed quasi-periodic oscillations (QPOs) in BH-XRBs  \citep[e.g.,][]{Fender-etal2004, Remillard-McClintock2006,Done-etal2007,Belloni2010,
2012A&A...542A..56N,2015ApJ...807..108I,Belloni-Motta2016,Ingram-Motta2019,2025arXiv250304011W}. Further, with semi-analytical general relativistic hydrodynamics, it has been shown that transonic solutions (including shocks) are possible not only around Schwarzschild/Kerr black holes \citep[e.g.,][]{1985A&A...148..176L,1993ApJ...412..254C,Chakrabarti1996,Gammie-Popham1998,Popham-Gammie1998,Chattopadhyay-Kumar2016,Dihingia-etal2018PhRvD,Dihingia2019LowIndex} but also around non-Kerr black holes \citep[e.g.,][]{Dihingia-etal2020PhRvD,Patra-eatl2022,Sen-etal2022,Uniyal-etal2024,Patra-etal2024}.

Recent investigations have shown the possibility of multi-transonic flows and the presence of standing or oscillating shocks in two-dimensional (2D) general relativistic hydrodynamic (GRHD) accretion flows \citep{2017MNRAS.472.4327S, Kim-etal2017, Kim-etal2019, Palit-etal2019}. In three-dimensional (3D) GRHD simulations, \cite{Sukova-etal2017} observed oscillating and expanding shocks. However, recent simulations by \cite{2023A&A...678A.141O} did not find clear evidence of such standing shocks. Still, \cite{2023A&A...678A.141O} reported some local density jumps caused by shocks, although these were not visible in the overall flow. Recent semi-analytical works by \cite{2024ApJ...971...28M} also suggest transonic shocks even in general relativistic magneto-hydrodynamic (GRMHD) flow. It is clear that such shocks are not limited to only hydrodynamic flow, and we need to explore more parameter space systematically. A recent study of GRMHD simulations of low-angular momentum flow around supermassive black holes has also suggested that such flow could produce centi-Hz QPOs and can be detected in polarization signatures \citep{2025ApJ...982L..21D}. It would be a new channel to seek the QPOs in supermassive BH accretion systems.

Based on the above discussion, it is clear that shocks in low-angular momentum flow have been studied in a wide range of frameworks, but not in GRMHD. Accordingly, this study focuses on exploring it, considering several axisymmetric (2D) simulations and a three-dimensional (3D) GRMHD simulation, and testing if such solutions exist in realistic environments around astrophysical black holes.
The pressure balance is essential in the formation of transonic shocks.
However, MAD flows are associated with very high magnetic pressure close to the black holes and are often subjected to eruption events \citep{2023ApJ...950...31K, 2023ApJ...946L..42K}, which are expected to disturb the inner accretion flow structure. Accordingly, while searching for transonic shocks, we restrict %ourselves 
to investigate in SANE regime. While doing so, we closely follow \citet{2024ApJ...967....4D}. We extend the investigation by setting a parameter (labeled as $\alpha$) that restricts outer accretion %\textcolor{orange}{to be closer} 
closer 
to the equatorial plane. % ($\alpha$).    
Moreover, we investigate the effects of BH spin to the low angular momentum accretion flow in the formation of transonic shocks.

%It is quite useful to comprehensively study the low-angular-momentum accretion in the SANE case. The search of the standing shock is especially interesting. \citet{2024ApJ...967....4D} focused on the low-angular-momentum accretion in the SANE case. In this paper, we follow the investigation provided by \citet{2024ApJ...967....4D}. We extend the investigation to the equatorial accretion by setting different densify distributions of the torus surrounding the central BH. We find that the standing shocks can be originated from the equatorial inflow in the SANE case. Moreover, we investigate the effects from different BH spin to the cases mentioned above for a certain accretion inflow. We speculate that the standing shocks might be related to some observational phenomena, such as short-timescale flaring, quasi-periodic oscillation (QPO), or quasi-periodic eruption (QPE) that are shown in X-ray binaries (XRBs) and active galactic nuclei (AGNs).

We use the Black Hole Accretion Code (BHAC, \citet{2017ComAC...4....1P, 2019A&A...629A..61O}) to solve GRMHD equations. Section 2 presents the 2D methodology to achieve the scientific goals. In Sections 3 and 4, we present the results in detail. 
We illustrate the 3D validation test in Section 5.
A short summary and further discussion are given in Section 6.
%provides a concise overview, including a discussion on the prevailing standing shock, the validation of the torus, and certain observational features in the central BH region.

\section{Methodology and Simulation Setup} \label{sec:method}

%\subsection{Simulation Setup}
This study investigates accretion properties in an axisymmetric (2D) and full three-dimensional (3D) framework, described using $r$, $\theta$, and $\phi$ coordinates. The simulation domain is extended up to $r=2500\,r_g$. 
The effective resolution is set to be $1024\times 512$ (with two levels of static mesh refinement), and the minimum grid size in the radial direction is 0.1739.
We employ a unit system defined by $G=M_{\rm{BH}}=c=1$, where $G$ is the gravitational constant, $M_{\rm{BH}}$ is the BH mass, and $c$ is the light speed. In this unit system, the distance and the time are in units of $r_g=GM_{\rm{BH}}/c^2$ and $t_g=GM_{\rm{BH}}/c^3$, respectively. The numerical simulations in this paper are carried out up to $t=15,000\,t_g$ with the minimum time step of $\Delta t=1.0\times 10^{-6}\,t_g$, which is good enough to reach a quasi-steady accretion close to the black hole.

To set initial conditions, we utilize the Fishbone-Moncrief (FM) torus \citep{1976ApJ...207..962F,2024ApJ...970..172U}, and modify 
the angular velocity to sub-Keperian value by introducing a fraction ${\cal F}=0.4$ (fixed for the current study) following \citet{2024ApJ...967....4D}. Moreover, for consistency, we fix the inner edge and the density maximum of the torus at $r_{\rm in}=5\,r_g$ and $r_{\rm max}=18.5\,r_g$, respectively. This setting of the FM torus provides enough mass to study quasi-steady low-angular momentum flow in computational time. In addition, we introduce an inclination parameter, $\alpha$, as $\rho\propto \rm{exp}[-(\alpha\cos\theta)^2]$ in the density distribution obtained from the FM torus solution, where $\rho$ is the density. This parameter restricts the accretion direction of the inflow far from the black hole. We then choose the inclination parameter $\alpha$ to be $5.0, 10.0, 30.0$, and $50.0$ in the simulation setup to investigate its effects on the accretion flow targeting a transonic shock. We consider that the gas pressure in the simulations follows the ideal gas equation of state, $p\propto \rho^\Gamma$, with $\Gamma=4/3$.
Angular momentum produces centrifugal pressure, which is crucial for pressure balance during shock formation. The rotation of the black hole can also produce similar effects due to frame dragging close to the black hole. To study this, we perform the simulations with different spin parameter numbers $a_\ast=0.80,0.90,$ and $0.94$, and study their impacts.
Finally, a poloidal magnetic loop is introduced as an azimuthal component of the vector potential, i.e., $A_\phi\propto \rho/\rho_{\rm max} - 0.1$. This sets the initial magnetic field configuration in the FM torus. Additionally, we add a 4\% random perturbation in the gas pressure of the torus to excite MRI. 

%\ym{YM: No information about numerical grids, solving coordinates.}

%We take the azimuthal component of the contravariant four velocity to be $u^\phi$. As this component related to the angular momentum, we can set a ratio to reduce the number of this component. Although a certain magnetic field configuration is given in the FM torus, a low $u^\phi$ number may yield a SANE case, because a relatively small accretion velocity can accumulate relatively small magnetic energy. Here, we put the parameter to be 0.40, and we can take this accretion mode to mimic the SANE-case accretion. 

%\ic{HERE I SUGGEST SOME RESTUCTUREING OF THE TEXT}

\section{Temporal Evolution Study}

\begin{figure*}
    \centering
    \includegraphics[width=0.49\textwidth]{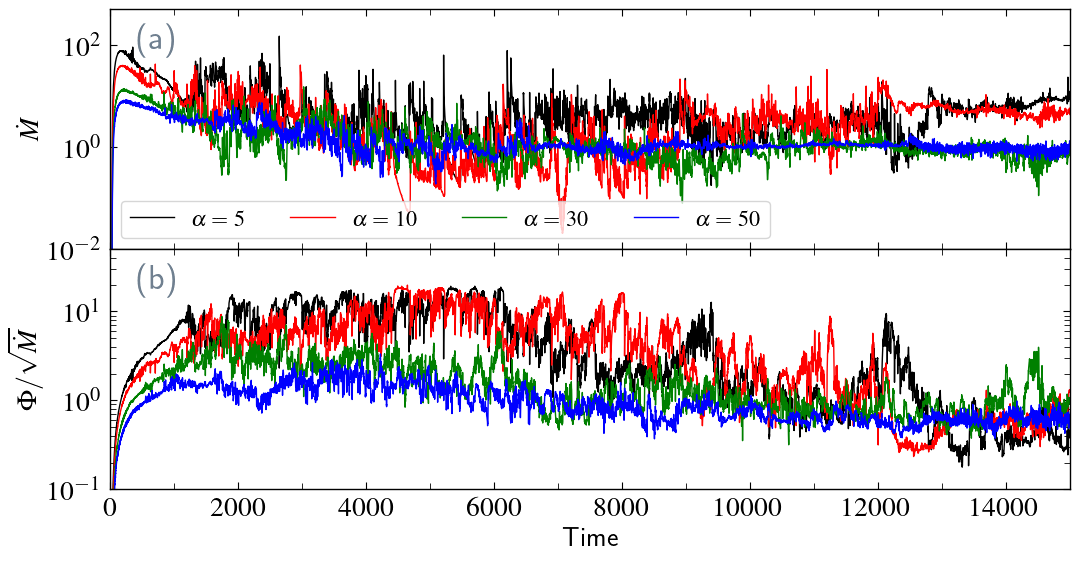}
    \includegraphics[width=0.49\textwidth]{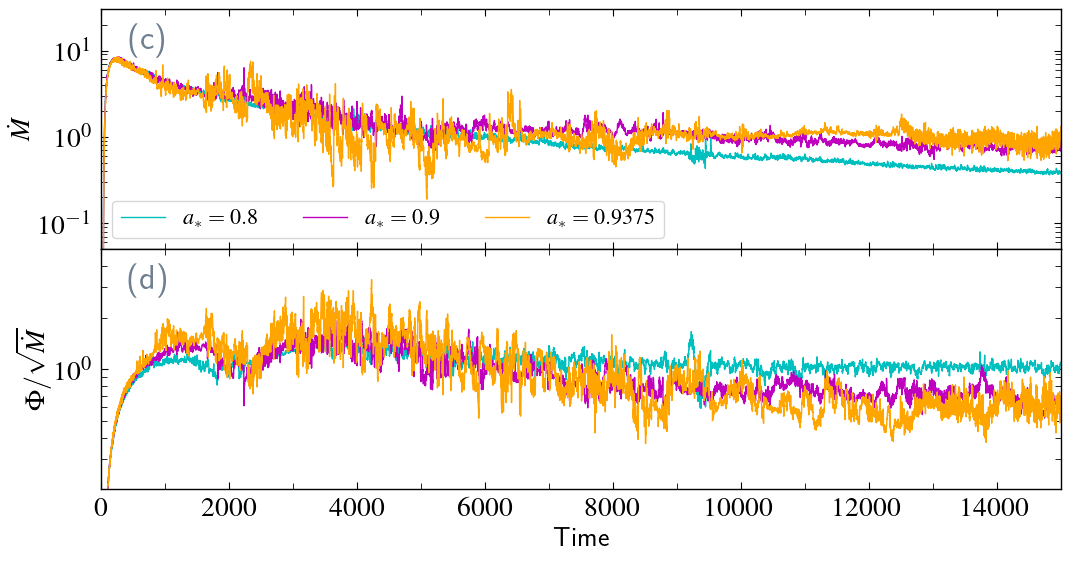}
    \caption{Time evolution of (a) accretion rate ($\dot{M}$) and (b) normalized magnetic flux ($\Phi/\sqrt{\dot{M}}$) at BH horizon. Left panels show variation for different $\alpha$ in the case of $a_\ast=0.94$, % \ym{YM: what is the BH spin value?}, 
    whereas the right panels show variations for different values of $a_\ast$ in the case of $\alpha=50$. %\ym{YM: It is better to change the labels to (c) and (d) for the right panels.}
    }
    \label{fig:01}
\end{figure*}

We illustrate the time evolution of the accretion flow properties in the low angular momentum case %accretion flow 
in this section. The time evolution behaviors of both the accretion rate $\dot{M}$ (in code units) and the normalized magnetic flux ($\phi/\sqrt{\dot{M}}$) at the BH horizon are shown in Figure~\ref{fig:01}. In the left panels of Figure~\ref{fig:01}, we present variations for different inclination parameters ($\alpha$) for fixed $a_\ast=0.94$, whereas in the right panels of Figure~\ref{fig:01}, we show variations for different spin parameters ($a_\ast$) for fixed $\alpha=50$. 
%%%%%%%% original sentence 
%Due to the low angular momentum nature of the flow, the flow is initially not in hydrostatic equilibrium. 
In our setup, the angular momentum profile is reduced to less than that of the marginally stable value of the respective black hole. Accordingly, it is not in hydrostatic rotational equilibrium as in the original FM torus solution. Therefore, the flow could fall onto the BHs without transferring angular momentum outwards. This explains why we initially observe an extremely high accretion rate. 
As infall time increases, making the MRI kick in, we observe an increase in the normalized magnetic flux with time. 
Subsequently, the accretion rate profiles show quasi-steady behavior until the end of our simulations. For the lower value of $\alpha$, we have more matter and magnetic flux around the BH. Accordingly, we see higher values of accretion rate and normalized magnetic flux at the horizon. 
However, after a longer simulation time ($t\gtrsim 10,000\,t_g$), MRI can not continuously grow the magnetic fields due to the 2D nature of the simulations. Therefore, its values drop to a similar level for all cases. In such cases, we obtain  
$\Phi/\sqrt{\dot{M}}\sim1$, which is well below MAD limits \citep{2011MNRAS.418L..79T}; this confirms that our accretion models are in the SANE regime. Nonetheless, the magnetic flux shows a higher variability magnitude for lower values of $\alpha$. 

We see three different BH spin cases of $a_\ast=$0.8, 0.9, and 0.94, respectively, in the right panel of Figure~\ref{fig:01}. The aim is to investigate how BH spin, via frame-dragging effects, influences low-angular-momentum accretion flows for a fixed ${\cal F} = 0.4$. As the BH spin parameter increases, the frame-dragging frequency near the horizon also increases, with the angular velocities at the horizon given by $\Omega_{\rm H} = a_*/2r_{\rm H}$ ($r_{\rm H}:$ location of the event horizon), resulting in values of $0.250$, $0.314$, and $0.350$ for $a_* = 0.8, 0.9,$ and $0.94$, respectively. This enhanced spacetime rotation increases the effective angular velocity of the inflowing plasma near the BH. As a result, the centrifugal support in the inner regions becomes stronger, which can help sustain more matter closer to the BH for a longer time by delaying the rapid inward plunging of matter. Consequently, we observe that the mass accretion rate remains relatively higher over time in the higher-spin cases, while in the lower-spin cases, the accretion rate decreases more quickly. Accordingly, we also see weaker normalized magnetic flux for higher-spinning BH cases as compared to lower-spinning cases. Note that the normalized magnetic flux can decrease even if the magnetic flux $\Phi$ remains constant or increases slowly due to the frozen-in condition. It provides a dimensionless measure to compare the dynamical importance of the magnetic field relative to the accretion or inflow across systems with different accretion rates.
%\textbf{Accordingly, as mentioned before, after a long simulation time, MRI does not continuously grow the magnetic fields in the 2D simulations, as this corresponds to the SANE mode in all the cases. However, the mass accretion accumulates as the simulation time increases. Thus, we also see weaker normalized magnetic flux for higher-spinning BH cases as compared to lower-spinning cases.}
We repeat that all the cases are in the SANE regime. In the MAD regime, due to significant magnetic pressure close to the black hole, a well-developed MRI may be necessary for accretion to happen. We plan to study such cases in the future.

\begin{figure*}[h!]
    \centering
    \includegraphics[scale=.3]{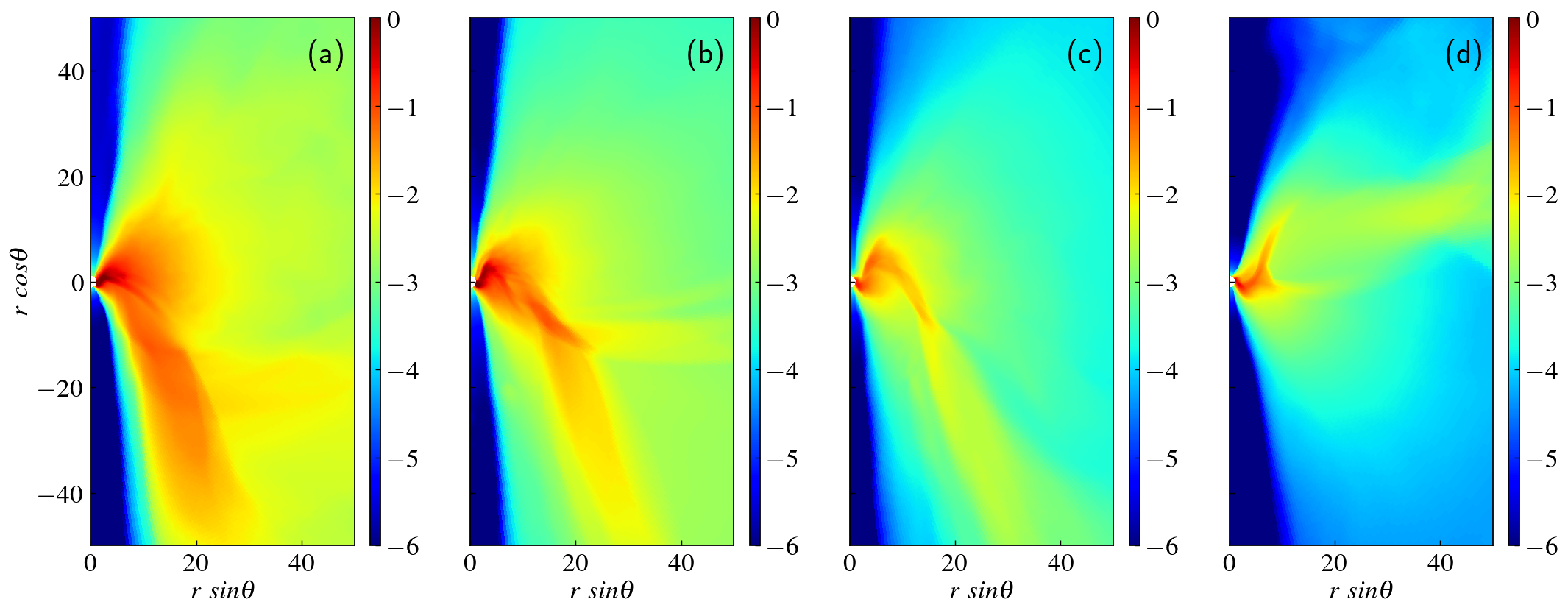}
    \includegraphics[scale=.3]{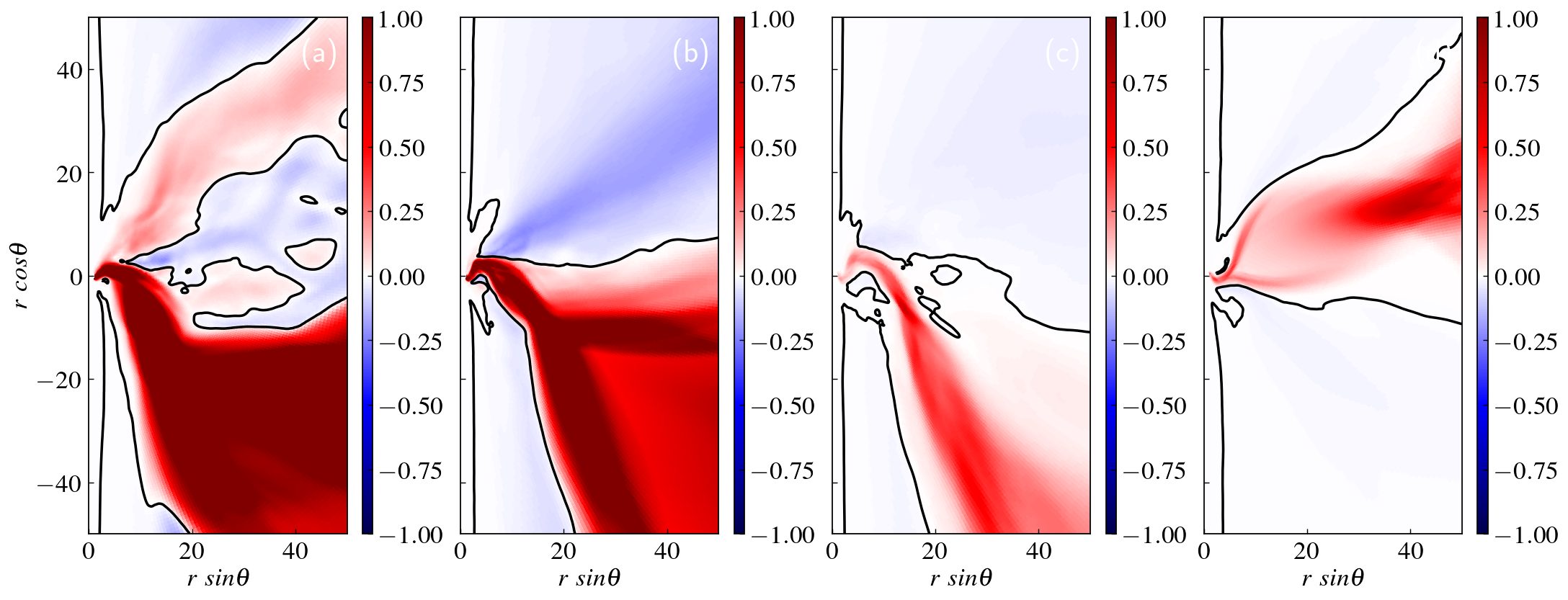} 
    \caption{Distribution of logarithmic density ({\it upper}) and mass accretion rate ({\it lower}) for different inclination parameter $\alpha=5$  (panel a), 10 (panel b), 30 (panel c), and 50 (panel d), respectively. We keep the BH spin to be $a_\ast=0.94$. The black contour indicates the boundary where the accretion rate is equal to zero. %\ym{YM: what is the black contour line?}
    }
    \label{fig:02}
\end{figure*}

\begin{figure*}[h!]
    \centering
    \includegraphics[scale=.3]{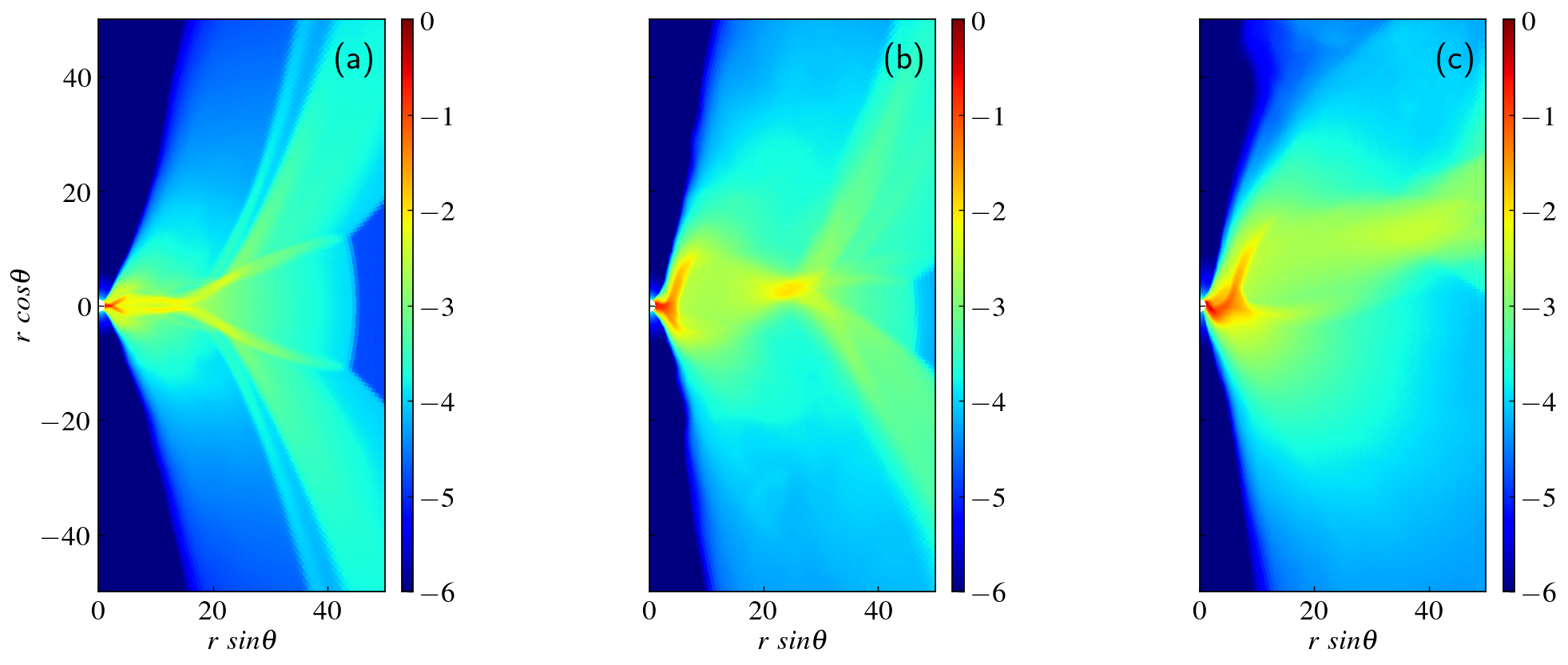}
    \includegraphics[scale=.3]{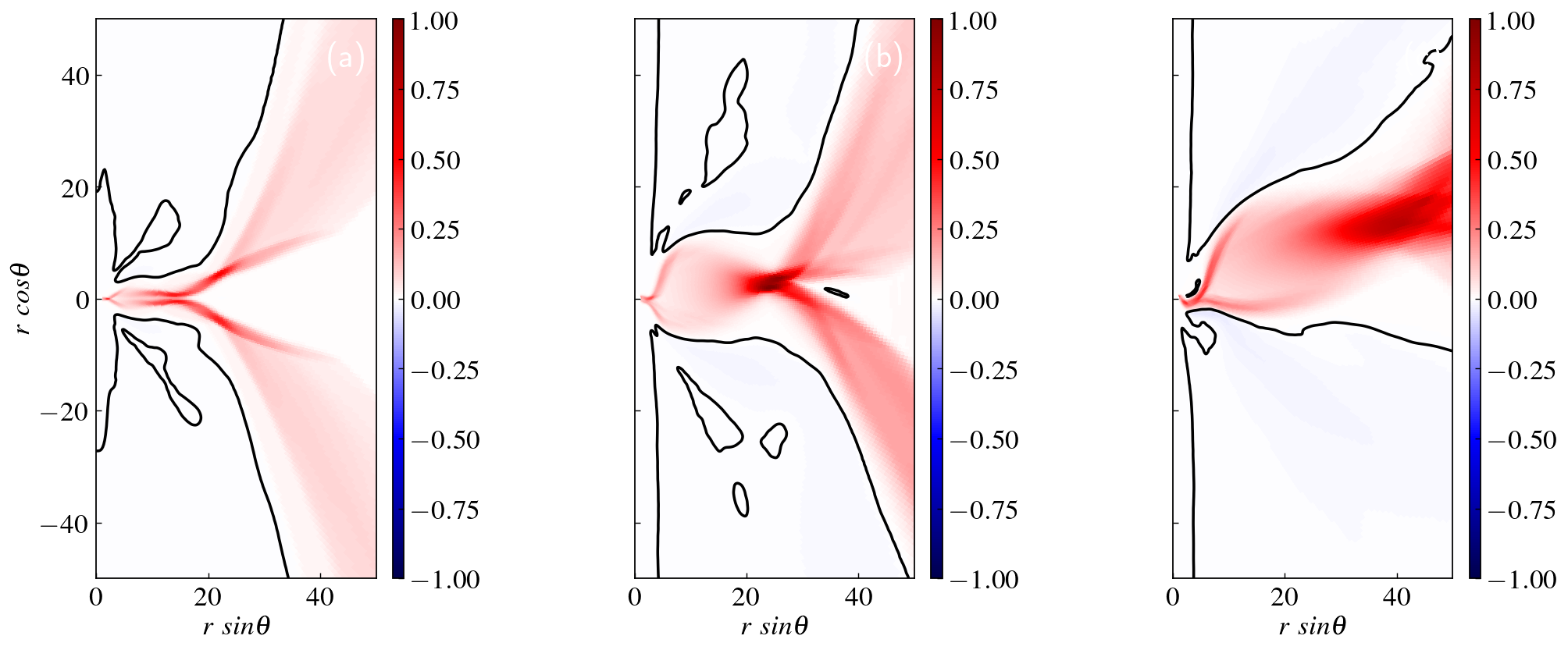} 
    \caption{Same as Fig~\ref{fig:02} but shown for different BH spin parameter $a_\ast$=0.8 (panel a), 0.9 (panel b), and 0.9375 (panel c), respectively.
    We keep the inclination  
    parameter to be $\alpha=50$.}
    \label{fig:03}
\end{figure*}

\begin{figure*}[ht!]
    \centering
\includegraphics[scale=.3]{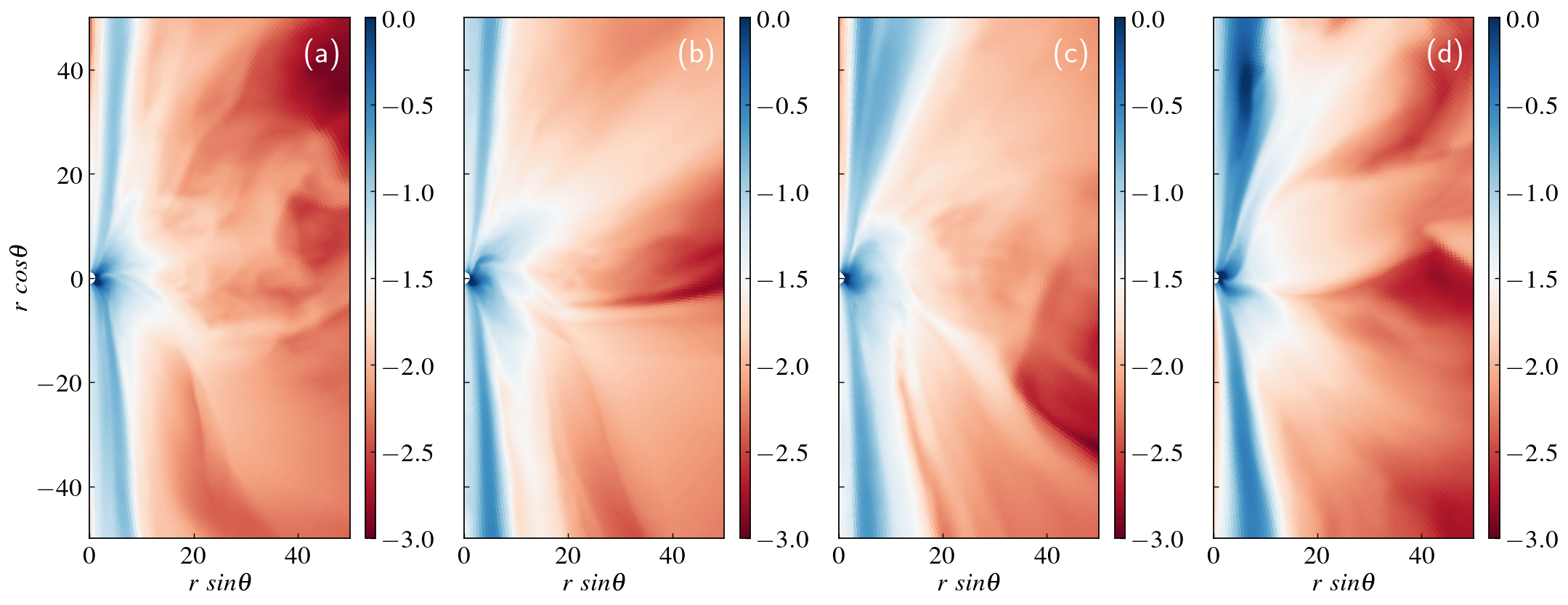}
\includegraphics[scale=.3]{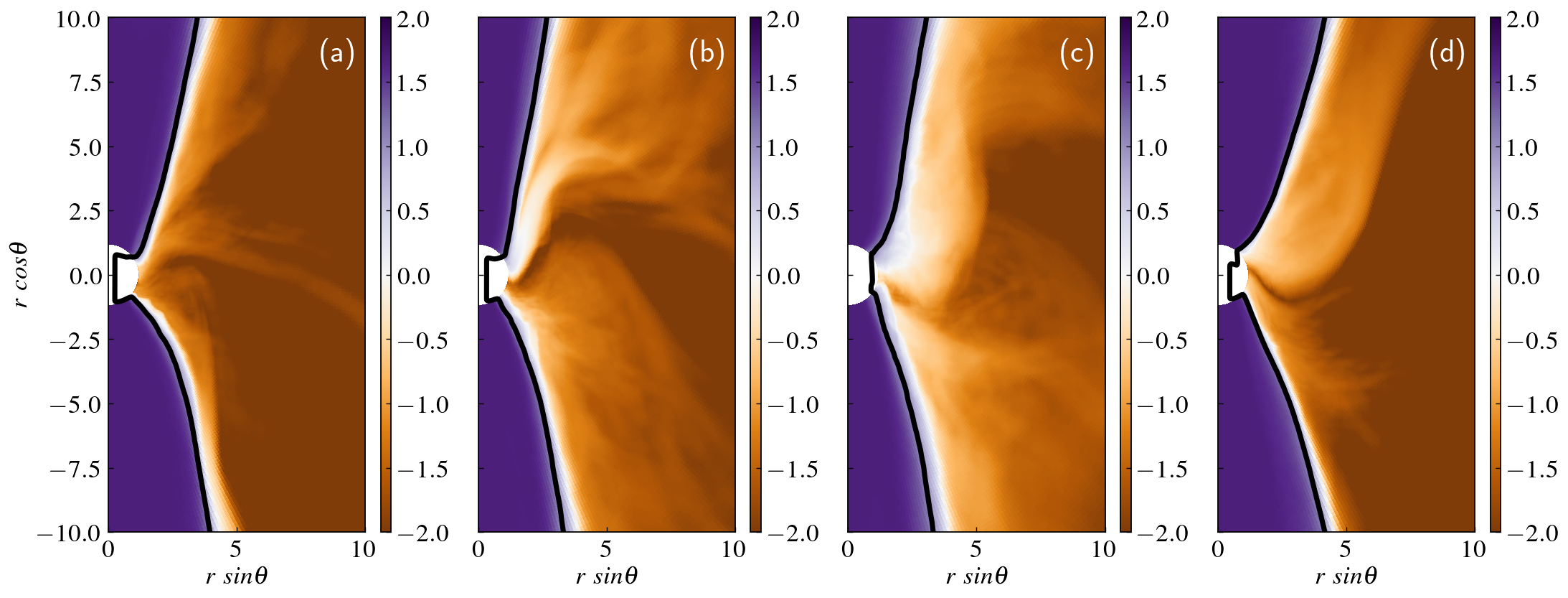}
    \caption{Same as Fig~\ref{fig:02} but shown for the distribution of Lorentz factor ($\log(\gamma-1)$, {\it upper}) and magnetization ($\log(\sigma)$ {\it lower}). The black contour indicates that the magnetization is equal to 1.}
    \label{fig:04}
\end{figure*}

\begin{figure*}[ht!]
    \centering
\includegraphics[scale=.3]{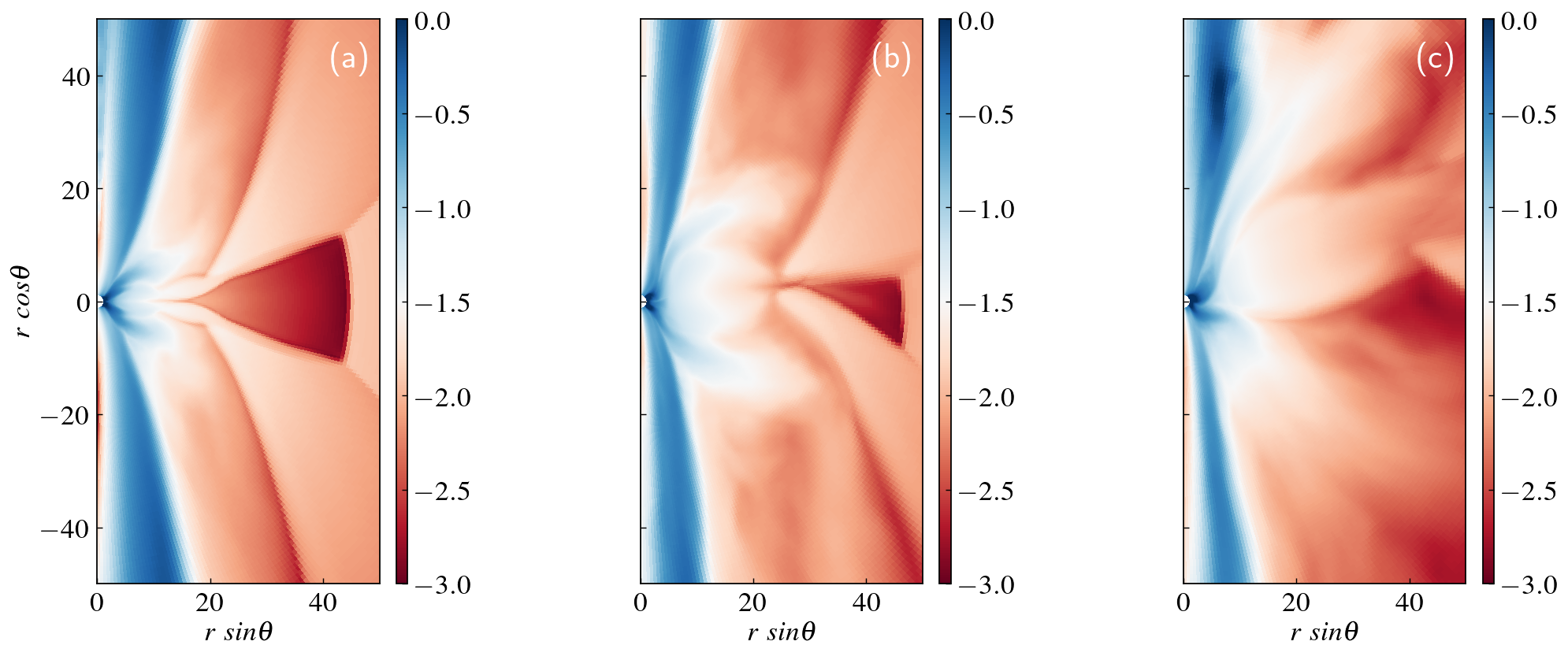}
\includegraphics[scale=.3]{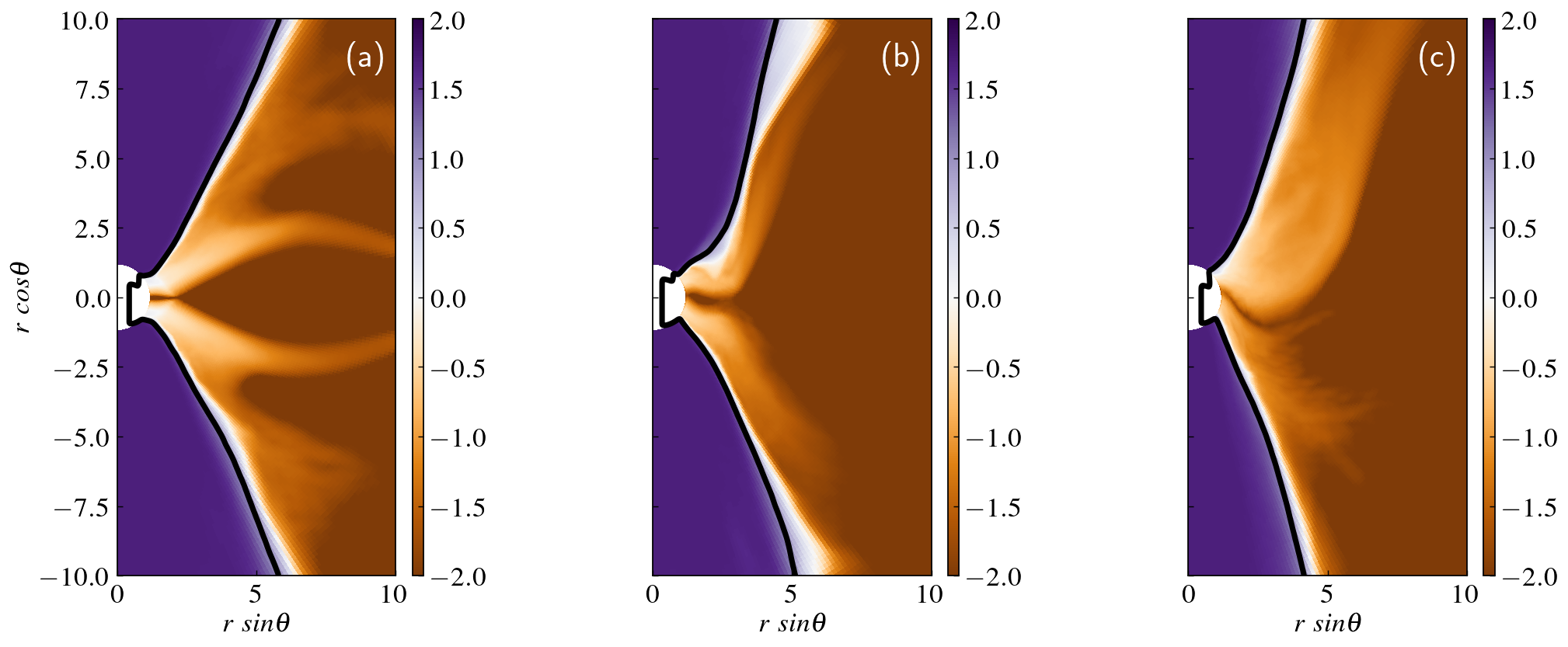}
    \caption{Same as Figure~\ref{fig:03} but shown for the distribution of Lorentz factor ($\log(\gamma-1)$, {\it upper}) and magnetization ($\log(\sigma)$, {\it lower}). }
    \label{fig:05}
\end{figure*}

%\clearpage

\section{Averaged flow behavior}

In this section, we illustrate the time-averaged flow properties of the accretion in the low angular momentum cases. In Figure~\ref{fig:02} and Figure~\ref{fig:03}, we show the distribution of time-averaged density and mass flux for different inclination and spin parameters, respectively. The averaging is performed over simulation time $t=12000-15000\,t_g$.

%\item Give some discussion on the general features of the density and mass flux distribution.

%\item Similar to the previous section, explain the behaviour changes with the parameters one at a time. Try to explain these changes physically. 

In Figure~\ref{fig:02}, the differences are clearly shown in the accretion flows around the equatorial plane from the time-averaged density distribution. 
In particular, the standing shock appears around $r\sin\theta\sim 10~r_g$ with an arc structure, as shown in the upper panel (d) of Figure~\ref{fig:02} (please follow Section~4.1 for more quantitative discussions).  
It is indicated that the standing shock exists because the accretion flows are more concentrated near the equatorial plane by setting a larger inclination parameter ($\alpha=50$). In other cases with smaller numbers of $\alpha$, we do not find any clear evidence of the presence of standing shocks. 
The distribution of the time-averaged mass flux is shown in the lower panels of Figure~\ref{fig:02}. When the inclination parameter becomes larger, the accretion flows (inflowing mass flux) are more concentrated on the equatorial plane.

The distribution of time-averaged density for different BH spin cases is shown in the upper panels of Figure~\ref{fig:03}. 
We can see the presence of the standing shock structure in the first two cases ($a_*=0.8, 0.9$) for %$r\sin\theta \sim 50\,r_{g}$ 
$r\sim 50\,r_{g}$ 
around the equatorial plane. More quantitative descriptions can be found in Section 4.1.
Interestingly, it seems that the position of the developed shock structure moves outward with increasing BH spin $a_\ast$.
The distribution of time-averaged accretion rate (mass flux) for each BH spin case is shown in the lower panels of Figure~\ref{fig:03}. It seems that a case of larger BH spin has a stronger accretion flow around the equatorial plane. 
As the BH spin becomes larger, the centrifugal force by frame-dragging effect becomes stronger. The standing shock structure is affected by this centrifugal effect. Thus, it can be shown in the outer position.

The distribution of time-averaged Lorentz factor and magnetization is shown in Figure~\ref{fig:04} (the effect on the different inclination parameter) and Figure~\ref {fig:05} (the effect on the different BH spin).
From Figure~\ref{fig:04}, it seems that the Lorentz factor in the jet funnel region is slightly higher as the inclination parameter $\alpha$ is increased. However, magnetization does not have a significant change in the different inclination parameters. 
It is indicated that the magnetized funnel region can always be formed in different inclination parameter cases when the BH spin keeps a high value. 
From the distribution of time-averaged Lorentz factor for the different BH spin cases seen in Figure~\ref{fig:05}, it is seen that some clear patterns indicate the standing shock structure in accretion flows.
The distribution of time-averaged magnetization for different BH spin cases shows magnetized funnel region is slightly increased as the BH spin becomes larger. 

\subsection{Standing shocks}
%\begin{itemize}
%    \item Figs: 1D shock analysis figures.

\begin{figure}[ht!]
    \centering
    \includegraphics[width=.95\columnwidth]{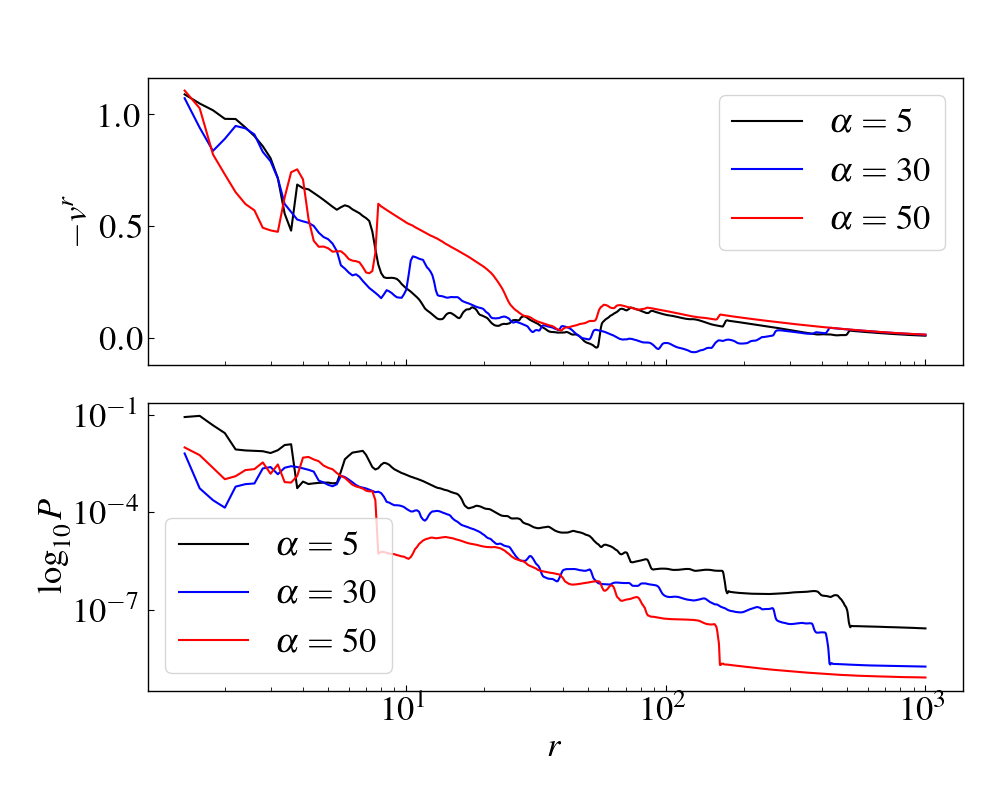}
    \includegraphics[width=.95\columnwidth]{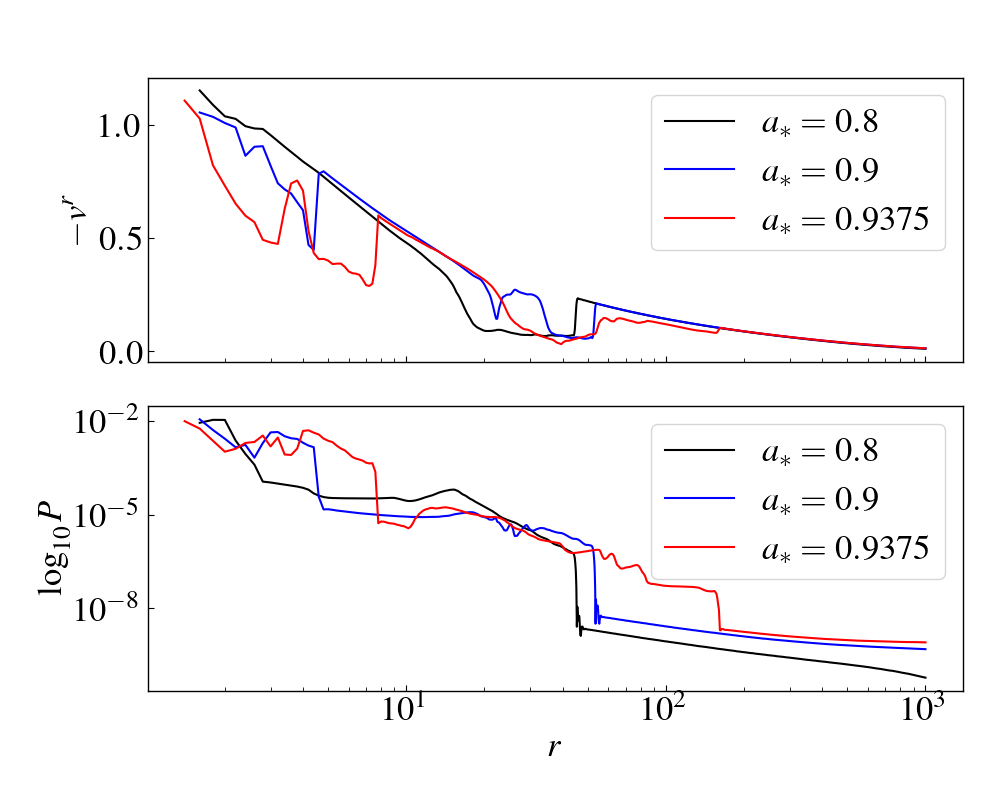}
    \caption{The radial velocity $v^r$ and the pressure $P$ as a function of $r$ at the time of $t=15,000~t_g$ for the standing shock identification. Upper panel: the case of $a_\ast=0.9375$; Lower panel: the case of $\alpha=50$.}
    \label{fig:06}
\end{figure}

\begin{figure}[ht!]
    \centering
    \includegraphics[width=.95\columnwidth]{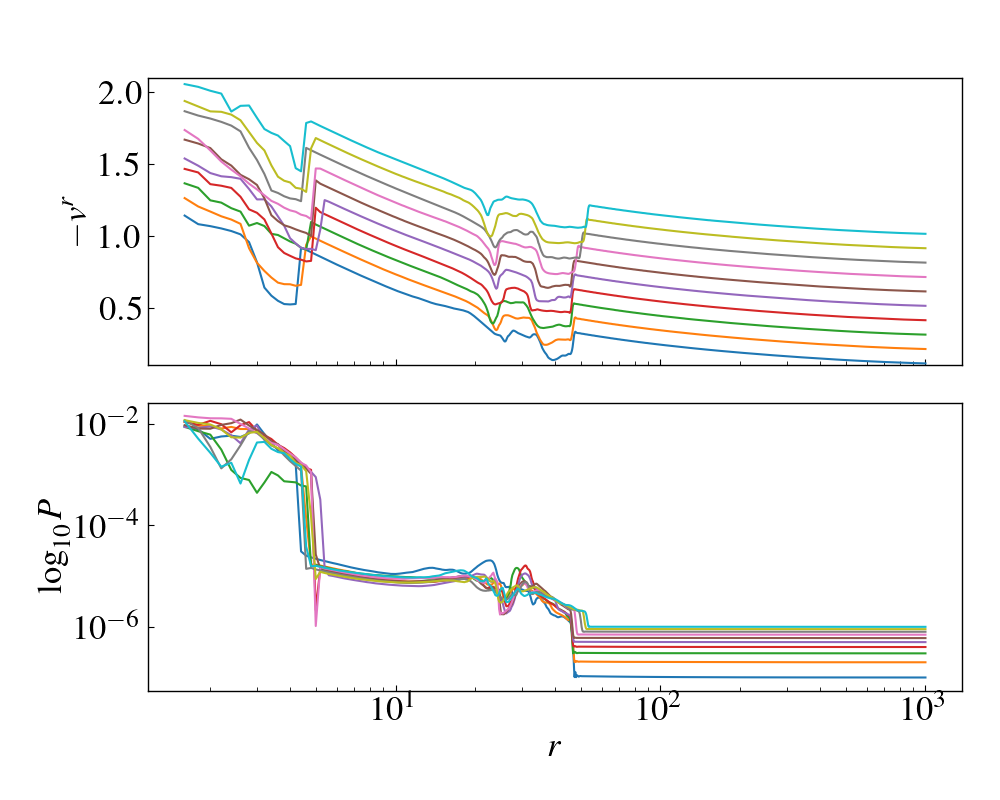}
    \caption{The standing shock identified in the case of $\alpha$=50 and $a_\ast=0.9$. The evolution time has the range of $14,000-15,000~t_g$ with the time interval of 1,000 $t_g$ (The lines are from bottom to top with the time increasing in each panel). Upper panel: radial velocity $v^r$ as a function of $r$; Lower panel: pressure $P$ as a function of $r$.}
    \label{fig:07}
\end{figure}
    
%We present the standing shock that is identified in our simulation in this subsection. 
From the distribution of the density in low angular momentum simulations, we obtain some hints of the presence of standing shocks. In order to fully understand the standing shock properties, we analyze the accretion flow structures along the equatorial plane.
%, one-dimensional (1D) shock analysis is presented. We illustrate the shock structure in Figure 6.

Figure~\ref{fig:06} shows the one-dimensional profile of radial velocity and the pressure along the equatorial plane at $t=15,000\,t_g$ %, the entropy, and the Mach number 
as a function of the radius in the cases of different inclination parameter $\alpha$ and different BH spin $a_\ast$. 
%As an example, the velocity and the pressure in different cases of $\alpha$ or $a_\ast$ at the time of 15,000 $t_g$ are given. 
The presence of the standing shock is clearly shown in all the panels of Figure~\ref{fig:06}. 
%\ym{YM: for me, it looks like multiple shocks seen in the 1D profile. Could you explain which one you focus on?}
%*****
%\textcolor{orange}{At least, in each case, two prominent standing shocks are clearly shown. In the meanwhile, some other small shock structures are also presented. The research on the shocked accretion flow has been developed since the time that the accretion was proposed. \citet{1985A&A...148..176L} found the transonic solution in relativistic accretion flow by an analytical way,
%and \citet{2006ApJ...645.1408T} obtained the fast ans slow magetosonic shocks in GRMHD accretion flow. 
%The standing shock was suggested in the hydrodynamical accretion flow \citep{1989ApJ...347..365C, 1997A&A...321..665L,2001ApJ...557..983D}. 
%\citet{2016ApJ...819..112L} suggested inner shock and outer shock in the ADAF framework. 
%Furthermore, dynamical instability of standing shock was proposed \citep{1994MNRAS.270..871N}.
%The instability of the standing shock may produce different modes \citep{2009ApJ...696.2026N, 2016ApJ...819..112L}. For a thin disk, it seems that the inner shock is unstable and the out shock is always stable \citep{2022MNRAS.512.5771H}. 
%Our results in this paper in corporation to the recent exploration of the standing shock by \citet{2023A&A...678A.141O} and \citet{2024ApJ...971...28M} provide further development on the research of the shocks in accretion flow.
%In our work, we clearly detect the standing shock signature in the case of the lower-angular momentum accretion. In particular, in our work,}
In particular, in our work,
it seems that we have a stronger shock signature (pressure jump becomes more significant) as the inclination parameter is larger. Similarly, when the BH spin is larger, the shock structure becomes clearer.
%it seems that the shock signature becomes stronger as the BH spin is stronger.
It is indicated that the standing shock is inclined to occur in the case of high-spin and strong equatorial accretion.     

It seems that two shocks are clearly identified in the case of highly-inclined accretion $\alpha=50$, while they are not clearly shown in the other two cases with low inclined accretion. Conversely, in the case of highly inclined accretion, we observe the two shocks across all BH spin cases. Moreover, it seems that the shock in the case of larger BH spin occurs at a bit larger distance from the central BH. It is because the centrifugal pressure (due to centrifugal force) increases with the spin of the black hole. Accordingly, shock fronts move far from the central objects as compared to low-spin black hole cases.

%\item First show the time evolution to justify that these shocks are more or less standing.
%\item Second, show how the shock location changes with the parameters and explain why.

We further illustrate the timing properties of the shock in our simulation. We take 10 slides from the time range of $14,000-15,000~t_g$ with the time interval of 1,000 $t_g$ and plot density and the pressure on the equatorial plane as a function of radii. 
We take the case of $\alpha=50$ and $a_\ast=0.9$.
The results are shown in Figure~\ref{fig:07}. The shock location appears consistently in all time slices around the same location, indicating temporal stability. Accordingly, the shock in our simulations can be identified as a standing shock. 
This is the first time that we show the formation of such standing shocks with GRMHD simulations. 
Earlier work of low angular momentum GRHD/GRMHD simulations \cite[e.g.,][]{2023A&A...678A.141O,2024ApJ...967....4D} did not report such features; our study is the first to identify stable standing shocks in weakly magnetized flows. This confirms that such structures are not only possible for idealized GRHD or semi-analytical models but also arise naturally in turbulent, weakly magnetized accretion flows around a rotating BH under suitable GRMHD conditions.

We note that two standing shocks, one is around $r=4-5~r_g$, the other is around $r=40-50~r_g$, are clearly identified in each case. 
%At least, in each case, two prominent standing shocks are clearly shown. 
Meanwhile, some other small shock structures are also presented. Research on the shocked accretion flow has been developed since the time that the accretion was proposed. \citet{1985A&A...148..176L} found the transonic solution in relativistic accretion flow by an analytical approach.
\citet{2006ApJ...645.1408T} obtained the trans-fast and slow magnetosonic solutions in stationary GRMHD accretion flows. 
The standing shock was proposed in the stationary hydrodynamical accretion flow \citep{1989ApJ...347..365C, 1997A&A...321..665L,2001ApJ...557..983D}. 
\citet{2016ApJ...819..112L} suggested the possibility of inner and outer shocks in the ADAF framework. 
Furthermore, the dynamical instability of a standing shock was studied by \citep{1994MNRAS.270..871N}.
The instability of the standing shock may excite different modes of oscillations \citep{2009ApJ...696.2026N, 2016ApJ...819..112L}. For a thin disk, it seems that the inner shock is unstable, but the outer shock is always stable \citep{2022MNRAS.512.5771H}. 
Our results in this paper, in corporation with the recent exploration of the standing shock by \citet{2023A&A...678A.141O} and \citet{2024ApJ...971...28M}, provide further development on the research of the shocks in GRMHD accretion flows.
In our work, we clearly detect the standing shock signature in the case of the lower-angular momentum accretion. %In particular, in our work,}
Compared to our work, as mentioned above, some former calculations in the early time  
used very idealized and restricted (1.5D) flows on the equatorial plane. 
Moreover, flow was considered to have certain energy and angular momentum, which was not a realistic expectation for accretion flow. In reality, accretion flows are turbulent and three-dimensional (we approximated it to be $2.5$D), leading to the formation of multiple standing shocks. These shocks can significantly impact the dynamics and observational signatures of accreting systems.
%\textbf{\ic{In the meanwhile, the stability of standing shock should be considered. For example, Bollimpalli et al. (2017) noticed the possibility of damping and instability of the shock in spherical accretion. This issue can be further examined in the future. DO WE REALLY NEED THIS SENTENCE, SPECIALLY THE CITATION. I DO NOT THIK IT IS NEEDED, REFEREE IS NOT ASKING FOR IT.}}

\section{3D GRMHD validation}
\begin{figure*}[ht!]
    \centering
    \includegraphics[width=0.49\textwidth]{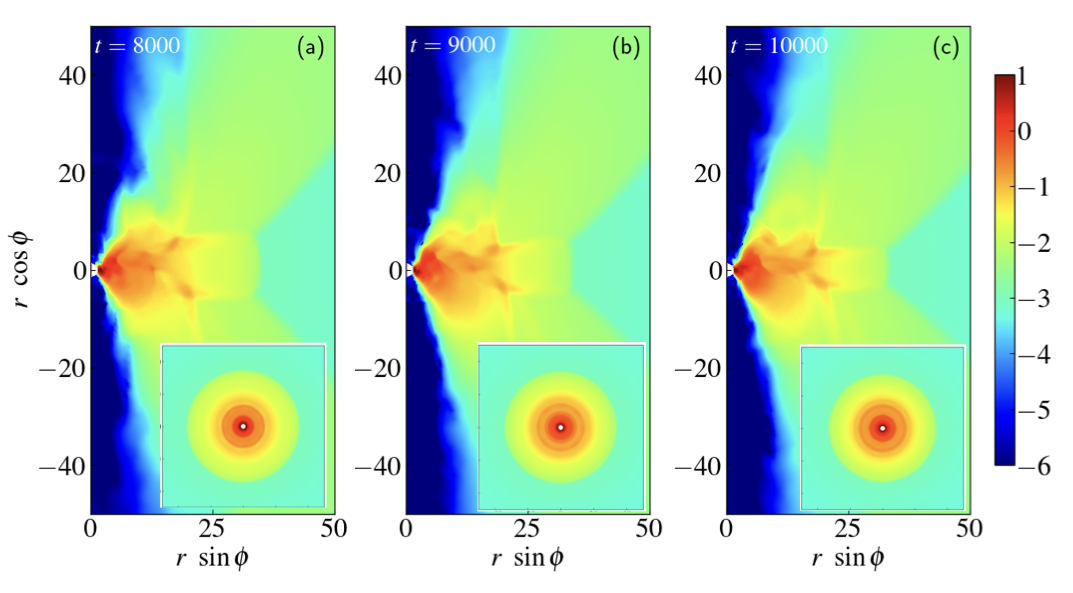}
    \includegraphics[width=0.49\textwidth]{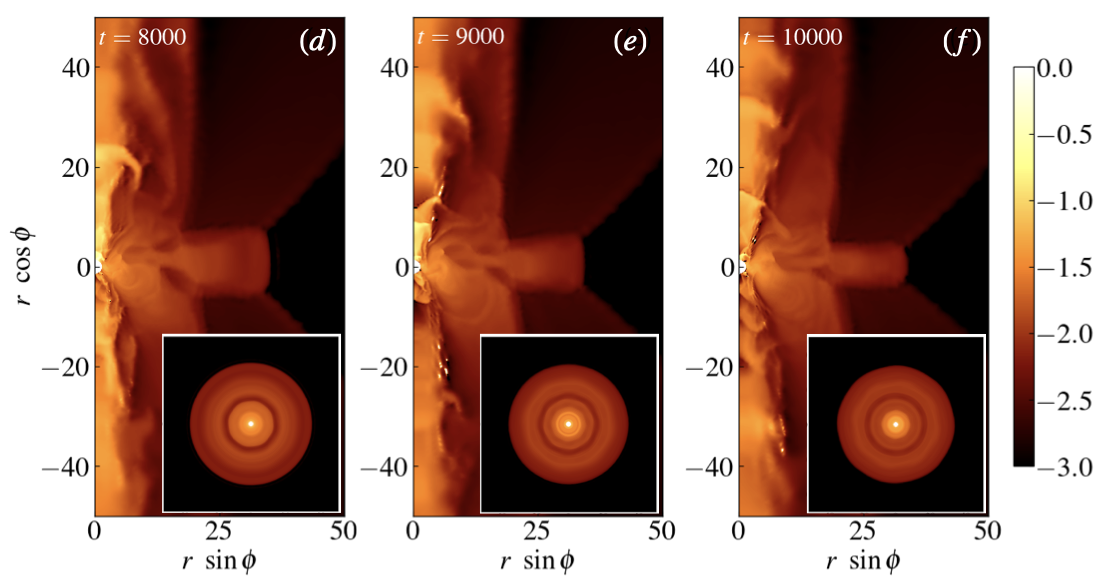}
    \caption{{\it Left:} Density $(\rho)$ and {\it Right:} Temperature ($p/\rho$) distribution on the poloidal plane $\phi=0$ for $3$D GRMHD simulations at different simulation times marked on each panel. The inset displays the distributions on the equatorial plane.}
    \label{fig:3d}
\end{figure*}

\begin{figure}[ht!]
    \centering
    \includegraphics[width=0.49\textwidth]{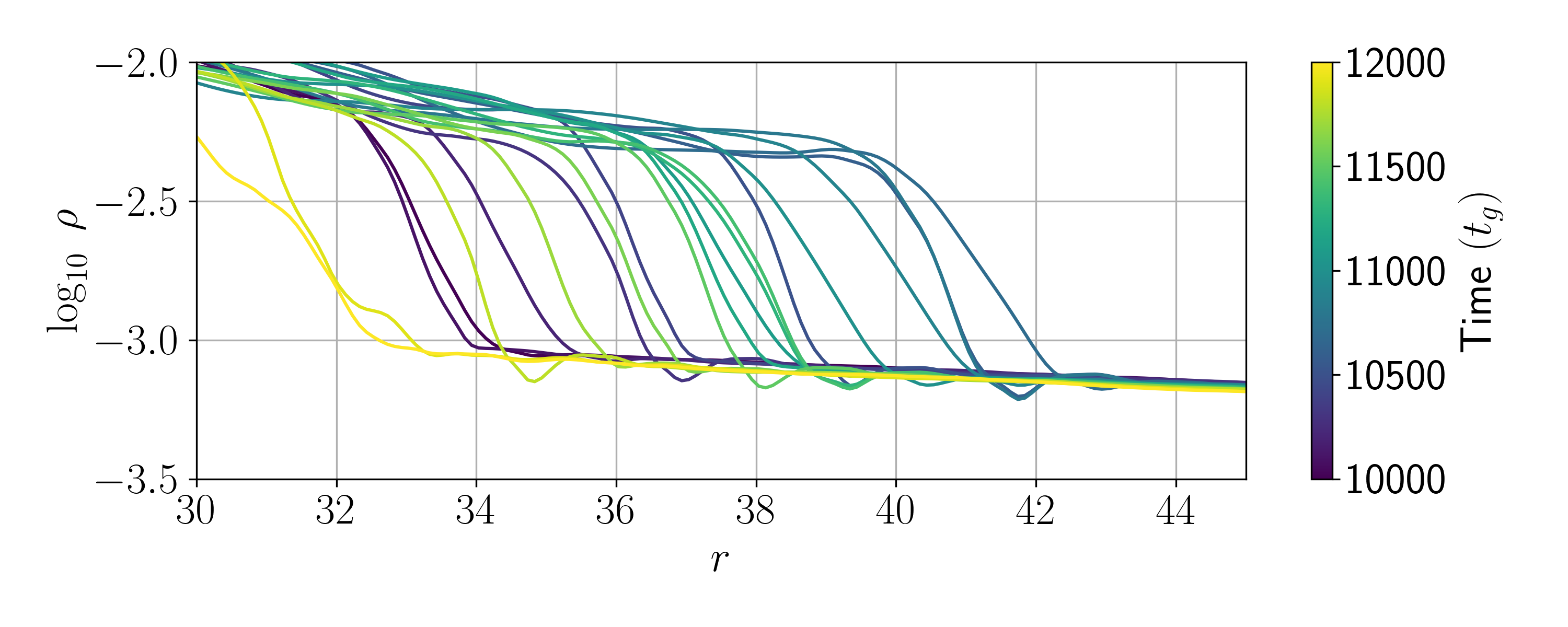}
    \caption{Radial logarithmic density profile $(\log_{\rm 10}\rho)$ on the equatorial plane around the shock location for different simulation time.}
    \label{fig:4}
\end{figure}

Throughout this study, we presented results based on our axisymmetric (2.5D) GRMHD simulations. In this section, we would like to demonstrate that the salient features of 2.5D GRMHD simulations remain consistent with full three-dimensional (3D) GRMHD simulations. For that, we perform 3D GRMHD simulation of a fiducial case that has $\alpha=10$, ${\cal F}=0.40$, and $a_*=0.9375$ with resolution, $256\times80\times64$. Since $\alpha=0$ case does not show formation of shock \citep[see][] {2024ApJ...967....4D}, we started with a lower inclination parameter $\alpha=10$. We find the formation of shock for a much lower value $\alpha$ than that of 2D cases. Since we successfully observed stable shock structures in this case, we did not perform the 3D simulations with higher $\alpha$ values due to the significantly higher computational cost of full 3D GRMHD runs. Nonetheless, we expect to see a similar shock structure for higher values, although the location of the shock surface may be different.
Figures~\ref{fig:3d}a-c and Figures~\ref{fig:3d}d-f show the 3D density and temperature ($p/\rho$) distribution on the poloidal plane of $\phi=0$ at simulation time, $t_s=8000$, $9000$, and $10,000$, respectively. 
In the inserted sub-panels, we show the same plots on the equatorial plane. All the panels show a sharp discontinuity in the quantities around $r\sim35\,r_g$, which can be identified as the shock location that we already observed in our 2.5D simulations. 
The shock location is quasi-steady with certain variations around its mean value. To show the shock locations at different simulation times ($t=10,000-12,000\,t_g$), in Figure~\ref{fig:4}, we show the radial logarithmic density profile $(\log_{\rm 10}\rho)$ on the equatorial plane by focusing around the shock formation region. The figure suggests that the shock location is not steady but oscillates with time (light and dark lines correspond to $t=10000\,t_g$ and $t=12000\,r_g$ are towards the left of the figure). A mean value of shock location is $r_{\rm s}\sim37\,r_g$, whereas the maximum and minimum shock locations are at $r_{\rm s, min}\sim32\,r_g$ and $r_{\rm s,max}\sim42\,r_g$, respectively. Therefore, the variation in shock locations with respect to the mean value is $\delta r_s \sim (r_{\mathrm{s,max}} - r_{\mathrm{s,min}})/2 \sim 5~r_g$.
Previous 2.5D hydrodynamic simulations also revealed such oscillations, which will be associated with the observed QPOs during outbursts of BH-XRBs 
\citep{2025ApJ...982L..21D}.

In the right panels of Figure~\ref{fig:3d}d-f, we see a sharp jump in the temperature distribution, which makes the post-shock region much hotter than that of the pre-shock region due to shock heating. From this result, we expect an abundance of hard X-ray emission %X-rays 
from the post-shock region. Furthermore, we observe that the post-shock region creates bipolar hot regions along the vertical direction, which could be associated with the post-shock corona (PSC, \cite{2015MNRAS.453.3414A}). Semi-analytical studies of low angular momentum flow have predicted such PSC for a long time \citep{2010MNRAS.401.2053D,2012MNRAS.421.1666S,2015MNRAS.453.3414A,2020MNRAS.497.2119M}, and accretion solutions involving post-shock flows were previously developed by \citet{1989MNRAS.240....7C,1996ApJ...471..237C} and \citet{2001A&A...379..683D}, along with several subsequent studies till date.
We confirm the formation of such PSCs with 3D GRMHD simulations. The location of PSCs helps the corona 
to effectively scatter soft photons, which might contribute to the high-energy tail of the observed X-ray 
spectra of BH-XRBs when they are in a hard state.
In the future, we would like to perform proper general-relativistic radiative transfer (GRRT) calculations to show this radiative signature explicitly.    

\section{Conclusions and Discussion} \label{sec:discussion}

In this work, we have conducted a systematic study of low angular momentum accretion flows using GRMHD simulations by the {\tt BHAC} code \citep{2017ComAC...4....1P, 2019A&A...629A..61O}, focusing on the conditions under which standing shocks can be formed and persist.
Unlike their high-angular-momentum counterparts, low-angular-momentum flows are characterized by quasi-spherical infall with limited centrifugal support, making them particularly sensitive to change in angular momentum, pressure balance, and black hole spin ($a_*$). The key conclusions of the study are outlined as follows:

\begin{itemize}
\item We verify that standing shocks form in low angular momentum accretion flows by using GRMHD simulations in the weakly-magnetized SANE regime. 

\item The presence of shocks is sensitive to both the inclination parameter ($\alpha$) and BH spin ($a_*$). We observed that the standing shocks are prominent in highly inclined (large $\alpha$) accretion flow cases.% (larger $\alpha$).

\item With the increase in the BH spin ($a_*$), the standing shock location shifts outward due to the stronger centrifugal effects of frame-dragging near the horizon. % due to frame-dragging.

\item %We saw that 
The shock locations remain quasi-steady over extended simulation time, indicating their stability and the persistently standing nature of the shock in the accretion flow.

\item 3D GRMHD simulations of low angular momentum flow support and validate the $2.5$D simulation results, showing the same accretion flow structures with shock transition and their long-term evolution.

\item The post-shock region is significantly denser and hotter, suggesting the formation of PSC, which is linked to the high-energy emission observed in BH-XRBs.
\end{itemize}

In summary, standing shocks can be formed in low angular momentum accretion flow for suitable combinations of flow and BH parameters. Several decades of semi-analytic calculations and hydrodynamic simulations of low angular momentum flow have predicted such solutions \cite[e.g.,][follow the ``Introduction" section for more references and discussions]{1989ApJ...347..365C, 2016ApJ...819..112L}. 
Our study confirms this with robust and cutting-edge GRMHD simulations. Due to the presence of a hot and dense PSC, such solutions are observationally viable for predicting the high-energy tail of the emission spectra of BH-XRBs \citep{2023MNRAS.525.4515H, 2025ApJ...980..251A}. 
We plan to study the role of PSC in generating high-energy spectra with proper GRRT calculations in the future. This will be able to link our simulations %directly 
to the observations in a direct way. 
%A recent study also suggests the low angular momentum flow around supermassive black holes could be detected in polarization signatures in terms of centi-Hz QPOs \citep{2025ApJ...982L..21D}. GHz band polarization observations of GRS~1915+105 have revealed QPO features \citep{2025arXiv250304011W}. It would be a new channel to seek the QPOs in supermassive BH accretion systems.

In this study, we have considered low angular momentum flow in the weakly magnetized SANE regime. However, the low angular momentum flow could also be in a magnetically dominated MAD regime \citep{2023ApJ...946L..42K}. In MAD, the inner accretion flow undergoes flux eruption events \citep[e.g.,][]{2011MNRAS.418L..79T, 2021MNRAS.502.2023P}. 
Then, we can not expect the formation of such standing shocks. On the contrary, we may see the formation of a
transient shock in the MAD regime, which needs to be studied in the future.

\section*{acknowledgments}
This work is supported by the National Key R\&D Program of China (No. 2023YFE0101200), the Natural Science Foundation of
China (No. 12273022, 12393813), CSST grant (CMS-CSST-2025-A07), 
the Shanghai Municipality Orientation Program of Basic
Research for International Scientists (No. 22JC1410600), and the Yunnan Revitalization Talent Support Program (YunLing Scholar Project). I.K.D. acknowledges the TDLI postdoctoral fellowship for financial support. 
S.N. is supported by JSPS Grant-in-Aid for Scientific Research (KAKENHI) (A) (No. JP25H00675).
We also acknowledge the support of the RIKEN-MOST program. %The simulations were performed on ...

\bibliography{main}{}
\bibliographystyle{aasjournal}

\clearpage

%\begin{figure}[htb!]
%    \centering
%    \includegraphics[scale=.6]{acc-mag-my-powerjet.png}
%    \caption{Upper panel: the effect of the equatorial accretion parameter $\alpha$ on the jet power $P_j$ at $r=50~r_g$. We keep the BH spin to be $a_\ast=0.9375$; Lower panel: the effect of the BH spin parameter $a_\ast$ on the jet power $P_j$ at $r=50~r_g$. We keep the equatorial accretion parameter to be $\alpha=50$.}
%    \label{fig:09}
%\end{figure}

\clearpage

\clearpage

\end{CJK*}
\end{document}